# Order parameter focalization and critical temperature enhancement in synthetic networks of superconducting islands

Francesco Romeo[1]

[1] Dipartimento di Fisica "E. R. Caianiello", Università di Salerno, I-84084 Fisciano (SA), Italy

E-mail: fromeo@sa.infn.it

**Abstract**

A generalization of the de Gennes-Alexander micronetworks theory is presented. In this framework, the phase transition of synthetic networks of superconducting islands is described by means of a Ginzburg-Landau approach adapted to the case of granular systems. The general implications of the theory are carefully explained. As a specific example, we demonstrate that star networks support the exponential localization of the order parameter accompanied by an enhancement of the critical temperature of the system. These findings contribute to clarify the physics of the phase transitions in synthetic networks of Josephson-coupled superconducting islands.

## 1. Introduction

In a pioneering work [1] R. Burioni and coworkers have demonstrated that spatial Bose-Einstein condensation can occour in dimension $d < 2$ when special discrete lattices are considered. The condensation of free bosons in these discrete structures is originated by an effective interaction induced by the network topology. In particular, it has been demonstrated that network nodes with higher connectivity act as localization centers for the boson density [2,3]. Following these authors, localization of free bosons on a graph can be studied by using suitable arrays of Josephson junctions.

This suggestion has been taken in a serious account and a relevant literature on this topic has been developed in recent times [4-7]. In particular arrays of Josephson junctions have been obtained and the Cooper pairs localization has been experimentally demonstrated. Surprisingly, sometimes these arrays show evidences of an enhancement of the system critical temperature compared to the critical temperature of the single superconducting islands. These experimental findings have been considered as the fingerprint of a Bose-Einstein condensation phenomenon in these structures [5]. Indeed, in virtue of the quasi-bosonic nature of the Cooper pairs, Josephson arrays should provide an analogue model for the condensation of bosonic atoms, the latter requiring extreme conditions to be experimentally realized.

However, Cooper pairs behaves like bosons only at low density, while deviations coming from the Pauli principle are expected to play a relevant role in the opposite regime. This observation suggests that a one-to-one identification between bosonic atoms on a network and Cooper pairs on a Josephson junctions array is not possible.

Interestingly, Josephson junctions arrays with special topology can be seen as coherent micronetworks mimicking the interstitial structure of the superconducting order parameter established at the phase boundary between the superconducting state and an insulating state. The latter situation is well known in oxide superconductors (e.g. $La_2CuO_4$), where the superconducting state emerges in close vicinity of a metal-insulator transition [8]. In the latter case, in vicinity of the superconducting phase transition, coherent regions nucleate inside an insulating matrix. These superconducting islands have a typical size of the order of the superconducting coherence length and are arranged in a random network in which the coherent regions are coupled by means of the Josephson effect. These networks present random topology and strength of the Josephson coupling. An intriguing property of these systems is that the connectivity among superconducting islands is much more relevant than the real dimensionality of the considered system. These arguments suggest that the connectivity properties of a given network determine whether a superconducting order can emerge starting from an insulating regime.

The above conclusions seem to be supported by the experimental evidences about the granular aluminium thin films [9-13]. These systems present insulating or superconducting properties depending on the growth conditions. A superconducting order can be induced in insulating systems by means of an electrical stress. Moreover, superconducting samples present a critical temperature enhancement compared to the aluminium thin film. All these observations suggest that in granular aluminium thin films superconducting grains form a network presenting random Josephson coupling and topology. Depending on the network topology, on the grains typical size and on the random distribution of the intergrain Josephson couplings, a superconducting order can emerge. Moreover, the network topology can be altered in a permanent way by applying an electrical stress.

From the theoretical side, the phase boundary between the superconducting and the insulating state has been first considered by P. de Gennes, who applied the Ginzburg-Landau (GL) theory to filamentary structures made of superconducting materials at micrometric level. The resulting theoretical approach, consisting in a GL theory on a graph, is known as de Gennes-Alexander micronetworks theory [14]. In its linearized version the de Gennes-Alexander theory allows the computation of the critical temperature of the system which depends on the external magnetic field and on the micronetwork topology.

In the present work, the de Gennes-Alexander micronetworks theory is adapted to the case of granular systems in which superconducting islands are coupled via the Josephson-type interaction. Using this theory, we demonstrate that a critical temperature enhancement emerges when specific network topologies are considered. The latter phenomenon is accompanied by the exponential localization of the superconducting order parameter. The exponential localization of the order parameter is peculiar to a discrete treatment and it cannot emerge in the context of the usual de Gennes-Alexander theory. These findings suggest a complementary interpretation of the Cooper pairs condensation reported in Josephson junctions arrays.

Moreover, it is expected that the order parameter focalization phenomenon, consisting in the exponential localization of the order parameter at the phase transition temperature, is not peculiar to superconducting systems. For this reason, it is argued that the above findings apply to generic phase transitions described by a Landau theory on a discrete lattice.

## 2. Methods

The de Gennes-Alexander micronetworks theory assumes that superconductivity nucleates inside an insulating matrix and thus takes an interstitial nature. In order to emulate this condition, a Ginzburg-Landau theory on a graph is formulated. The graph structure implies that several one-dimensional branches are connected to form a network. Solutions of the branch order parameters are constrained to be continuous and have to ensure the current density conservation at the network vertices. The above conditions provide appropriate boundary conditions to obtain the transition temperature and the order parameter profile at the transition temperature. The free energy of a generic branch is written in the form:

$$\mathcal{F} = \int dx \{\mathcal{C}|\partial_x \psi(x)|^2 + \mathcal{A}(T)|\psi(x)|^2 + \mathcal{B}|\psi(x)|^4\}, \quad (1)$$

where $\psi(x)$ represents the branch order parameter written in terms of the curvilinear coordinate $x$. As usual within the GL approach, the parameters $\mathcal{C}$ and $\mathcal{B}$ are positive and temperature independent quantities, while the sign of the coefficient $\mathcal{A}(T) = \alpha(T - T_c)$ changes at the transition temperature $T_c$. Here $T_c$ represents the bulk critical temperature of the superconducting material used.

In order to treat granular systems, we have to rewrite the branch free energy in terms of the order parameter $\psi_n$ of a generic island labelled by the discrete index $n$. Reasoning in the same spirit of the Lawrence-Doniach model, the following substitutions can be made in Equation (1):

$$\begin{aligned} \mathcal{C}|\partial_x \psi(x)|^2 &\to \mathcal{D}|\psi_{n+1} - \psi_n|^2 \\ \psi(x) &\to \psi_n \\ \int dx(\ldots) &\to \sum_n (\ldots) \end{aligned} \quad (2)$$

Observing that

$$|\psi_{n+1} - \psi_n|^2 = |\psi_{n+1}|^2 + |\psi_n|^2 - (\psi_n^* \psi_{n+1} + c.c.), \quad (3)$$

one immediately gets the following expression for the branch free energy:

$$\mathcal{F} = \sum_n \{(2\mathcal{D} + \mathcal{A}(T))|\psi_n|^2 - \mathcal{D}(\psi_n^* \psi_{n+1} + c.c.) + \mathcal{B}|\psi_n|^4\}. \quad (4)$$

Equation (4) represents the branch free energy of a granular system in which a Josephson-like coupling $-\mathcal{D}(\psi_n^* \psi_{n+1} + c.c.)$ between adjacent islands can be recognized.

Once the mathematical structure of the free energy of a single branch has been understood, the free energy of an arbitrary network can be presented in the following form:

$$\mathcal{F} = \sum_{ij} \psi_i^* H_{ij} \psi_j + \sum_i \mathcal{A}(T)|\psi_i|^2 + \sum_i \mathcal{B}|\psi_i|^4. \quad (5)$$

In writing Equation (5), we have explicitly assumed identical islands parameters. Moreover, to minimize the model complexity, the conditioning effect of an external magnetic field is not included in the model. The Hermitian matrix $H_{ij}$ is directly related to the network topology. In particular, $H_{ii} =$ 2𝒟, while $H_{ij}$ is equal to $-𝒟$ if the $i$-th and the $j$-th island are linked. $H_{ij} = 0$ in the remaining cases. Free energy in Equation (5) depends on the system temperature via the coefficient $𝒜(T)$. At the phase transition temperature, quartic terms in the islands order parameters can be neglected and the quadratic free energy

$$\mathcal{F}_\ell \approx \sum_{ij} \psi_i^* H_{ij} \psi_j + \sum_i 𝒜(T) |\psi_i|^2 \qquad (6)$$

can be considered to determine the phase boundary. Extremal condition for the free energy in Equation (6) can be written as follows:

$$\frac{\partial \mathcal{F}_\ell}{\partial \psi_k^*} = \sum_j H_{kj} \psi_j + 𝒜(T) \psi_k = 0. \qquad (7)$$

Previous equations can be presented in matrix form:

$$H\psi = -𝒜(T)\psi, \qquad (8)$$

where $\psi = (\psi_1, \cdots, \psi_N)^t$ collects all the islands order parameters. Equation (8) is solved when $\psi$ is an eigenvector of $H$. Let us introduce the eigenvectors $\phi^{(n)}$ of $H$ via the equation:

$$H\phi^{(n)} = \epsilon^{(n)} \phi^{(n)}, \qquad (9)$$

with $\epsilon^{(n)}$ the corresponding eigenvalues. Comparing Equation (8) and (9) we get $\epsilon^{(n)} = -𝒜(T)$. Thus $\phi^{(n)}$ is solution of the Equation (8) only when $T = T_c^{(n)}$, being $T_c^{(n)}$ given by:

$$T_c^{(n)} = T_c \left(1 - \frac{\epsilon^{(n)}}{\alpha T_c}\right). \qquad (10)$$

The transition temperature of the network $\mathcal{T}_c$ is given by

$$\mathcal{T}_c = max\{T_c^{(n)}\},$$

the latter condition being realized when the eigenvector $\varphi$ with the lowest eigenvalue $\epsilon_0 = min\{\epsilon^{(n)}\}$ is considered. Under this condition, the network critical temperature is given by

$$\mathcal{T}_c = T_c \left(1 - \frac{\epsilon_0}{\alpha T_c}\right), \qquad (11)$$

while the order parameter profile at the transition is proportional to $\varphi$. Equation (11) shows that the network transition temperature $\mathcal{T}_c$, which is a function of the bulk transition temperature $T_c$, depends on the network topology via the eigenvalue $\epsilon_0$. In this way, the value of the transition temperature $\mathcal{T}_c$ is an emergent property of the network topology. Interestingly, depending on the value of $\epsilon_0$, $\mathcal{T}_c$ can be either greater or lower than the bulk transition temperature. The order parameter profile can be determined as follows.

As stated before, the order parameter profile at the phase transition is proportional to $\varphi$. Thus, let us denote the order parameter profile as $\phi = a\,\varphi$, with $a$ a proportionality factor to be determined. To fix the proportionality factor, we have to minimize $\mathcal{F}[\phi]$ with respect to $a$ by retaining quartic terms in the free energy. In doing this, it is assumed that $\phi$, which has been obtained by neglecting quartic terms, does not differ too much from the actual order parameter profile at the phase transition. Explicit computation shows that the free energy takes the form $\mathcal{F}[\phi] = u\,a^2 + v\,a^4$, where we have introduced the notation:

$$\begin{array}{l} u = \sum_i \alpha(T - \mathcal{T}_c) |\varphi_i|^2 \\ v = \sum_i \mathcal{B} |\varphi_i|^4 \end{array}. \qquad (12)$$

A close inspection of Equations (12) shows that a non-trivial solution (i.e., $a \neq 0$) can be obtained only for $T < \mathcal{T}_c$. In that case, the extremal condition $\partial_a \mathcal{F}[\phi] = 0$ leads to the relation:

$$a = \Delta \sqrt{\frac{\mathcal{T}_c}{T_c}\left(1 - \frac{T}{\mathcal{T}_c}\right) \frac{\sum_i |\varphi_i|^2}{\sum_i |\varphi_i|^4}} \qquad (13)$$
$$\Delta = \sqrt{\frac{\alpha T_c}{2\mathcal{B}}}$$

where $\Delta$ represents the zero-temperature extrapolation of the bulk superconducting gap.

## 3. Results

In the previous section, we have provided a concise description of the theoretical formulation of a discretized Ginzburg-Landau theory on a graph. Before discussing specific cases, we provide a perturbative argument showing that highly connected islands tend to develop an enhanced order parameter.

Let us write Equation (5) in the following dimensionless form:

$$F = \eta \sum_{ij} f_i^* h_{ij} f_j - \sum_i |f_i|^2 + \sum_i \frac{|f_i|^4}{2}, \qquad (14)$$

where the zero-temperature limit is assumed. Dimensionless quantities have been defined according to the following relations:

$$\begin{array}{l} \mathcal{F} = \frac{(\alpha T_c)^2}{2\mathcal{B}} F \\ \psi_j = \Delta f_j \\ \eta = \frac{𝒟}{\alpha T_c} \\ H_{ij} = 𝒟\,h_{ij} \end{array}.$$

The free energy minimization condition requires that

$$\frac{\partial F}{\partial f_k^*} = \eta \sum_j h_{kj} f_j - f_k + |f_k|^2 f_k = 0. \qquad (15)$$

Assuming $\eta \ll 1$ a small parameter of the theory, Equation (15) can be solved by using a perturbation theory. Using the trial solution $f_k \approx 1 + \eta g_k$, we obtain

$$g_k = -\frac{1}{2}\sum_j h_{kj}.$$

Going back to dimensional variables, we get

$$\psi_k = \Delta f_k \approx \Delta \left\{1 - \eta + \frac{\eta}{2}\mathcal{N}_k\right\}, \quad (16)$$
$$\mathcal{N}_k = -\sum_{j \neq k} h_{kj}$$

where we have introduced the quantity $\mathcal{N}_k$ counting the Josephson couplings of the $k$-th island. In deriving Equation (16) we have explicitly used the property $h_{kk} = 2$. For a single infinite branch, Equation (16) implies that $\mathcal{N}_k = 2$ and $\psi_k = \Delta$, providing a spatially uniform superconducting solution. When the network is obtained by joining several branches, Equation (16) predicts an enhanced order parameter at the junction points of three or more branches, since $\mathcal{N}_k > 2$ at those points. Despite these speculations are based on a perturbative approach, numerical computation confirms the above conclusions.

In the following, relevant applications of the theory will be discussed.

### 3.1 Linear chain of islands

Let us consider a linear chain of $N$ islands. The free energy of this system can be written as done in Equation (4). Neglecting quartic terms, the extremal condition for the order parameter can be written in the form:

$$2\mathcal{D}\,\psi_k - \mathcal{D}(\psi_{k+1} + \psi_{k-1}) = -\mathcal{A}(T)\,\psi_k,$$

which is a special case of Equation (7). Following the general procedure outlined before, we obtain:

$$\psi_k = a\,\sin\left(\frac{\pi k}{N+1}\right)$$
$$\mathcal{T}_c = T_c\left\{1 - \frac{2\mathcal{D}}{\alpha T_c}\left[1 - \cos\left(\frac{\pi}{N+1}\right)\right]\right\}, \quad (17)$$

where the boundary condition $\psi_0 = \psi_{N+1} = 0$ has been used. Equations (17) provide the order parameter profile at the phase transition and the transition temperature $\mathcal{T}_c$. We explicitly notice that the branch critical temperature $\mathcal{T}_c$ is lower than $T_c$ and depends on the system size. In the thermodynamic limit (i.e., $N \to \infty$) we observe that $\mathcal{T}_c \to T_c$.

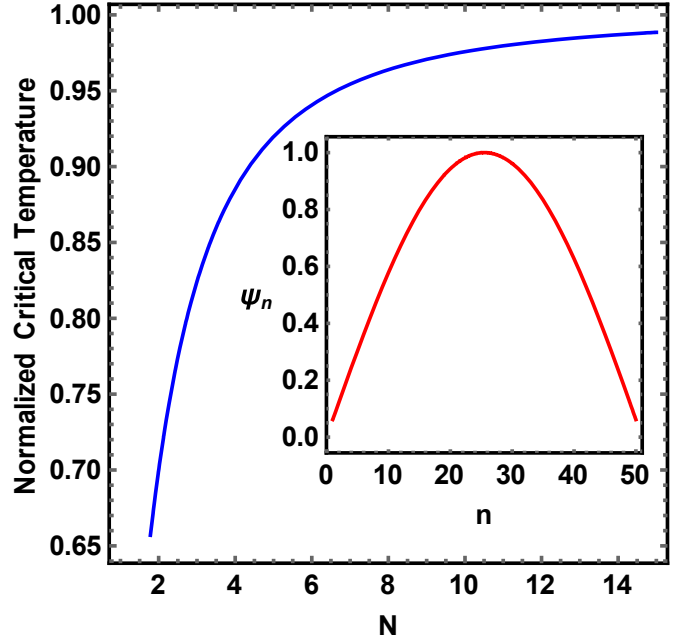

**Figure 1**. Normalized critical temperature $\mathcal{T}_c/T_c$ as a function of the system size $N$ obtained by using Equation (17) with $\eta = 0.3$. The inset shows the order parameter profile in arbitrary units at the phase transition for a system size $N = 50$.

The latter general observations are confirmed by direct computation whose results are shown in Fig. 1. In that figure a typical size-induced critical temperature reduction is observed, the latter being a well-known feature in thin films [15, 16].

Interestingly, as noticed in Ref. [15], the choice of the boundary conditions is not marginal and the chain critical temperature $\mathcal{T}_c$ depends on this choice. In deriving the above results absorbing boundary conditions (i.e., $\psi_0 = \psi_{N+1} = 0$) have been used. However, the alternative boundary conditions,

$$\psi_{N+1} = \beta_r\,\psi_N$$
$$\psi_0 = \beta_l\,\psi_1$$

with $\beta_{r,l} < 1$, could also be used. The latter are reminiscent of the de Gennes boundary conditions of a continuous theory, which take the form [17]:

$$\left(\frac{d\psi(x)}{dx}\right)_{x=L} = -\lambda_r\,\psi(x = L)$$
$$\left(\frac{d\psi(x)}{dx}\right)_{x=0} = \lambda_l\,\psi(x = 0)$$

The boundary conditions choice is not arbitrary and depends on the physical constraints that the environment induces on the system under investigation.

A further observation deserving to be discussed is the following. In the thermodynamic limit the system coherence length is a divergent quantity at the phase transition temperature $T_c$. It is now possible to show that the coherence length of the finite size system remains finite at the system transition temperature. Indeed, the dimensionless coherence length $\xi(\mathcal{T}_c)$ at the phase transition temperature $\mathcal{T}_c$ is given by the relation:

$$\frac{1}{\xi^2(\mathcal{T}_c)} = -\frac{\mathcal{A}(\mathcal{T}_c)}{\mathcal{D}}. \tag{18}$$

Equation (18) can be written in the following form:

$$\xi(\mathcal{T}_c) = \frac{\xi_0}{\sqrt{1-\frac{\mathcal{T}_c}{T_c}}} \\ \xi_0 = \sqrt{\frac{\mathcal{D}}{\alpha T_c}} \tag{19}$$

Using the second relation of Equation (17) in Equation (19) we obtain:

$$\xi(\mathcal{T}_c) = \frac{1}{\sqrt{2\left[1-\cos\left(\frac{\pi}{N+1}\right)\right]}} \approx \frac{N+1}{\pi}, \tag{20}$$

the latter being a good approximation when $N \geq 3$. Equation (20) shows that $\xi(\mathcal{T}_c)$ is a finite quantity at the phase transition corresponding to a fraction of the system size. The correct thermodynamic limit is recovered when $N \to \infty$. In view of these equations, the second relation in Equation (17) takes the approximate form:

$$\mathcal{T}_c \approx T_c \left[1 - \left(\frac{\xi_0}{\xi(\mathcal{T}_c)}\right)^2\right],$$

showing that the critical temperature suppression starts to operate when $\xi(\mathcal{T}_c)$ is comparable with $\xi_0$.

### 3.2 Linear chain coupled to a single extra site

So far, we have discussed the phase transition of a linear chain of superconducting islands. We have also noticed that islands showing higher connectivity present an increased order parameter. In order to show this effect, we add to the linear chain discussed before a single extra site. In this way, the system shown in Figure 2 is obtained. Let us denote with $\chi$ the order parameter of the extra site. The extra site is linked to the linear chain of size $2N + 1$ via a Josephson coupling contribution to the system free energy given by $-\Gamma(\chi^*\phi + c.c.)$, where $\phi$ represents the order parameter of the island belonging to the chain and linked to the extra site. The system variables $x_n$ and $y_n$ represent the order parameters of the linear chain above and below the special island labelled by $\phi$.

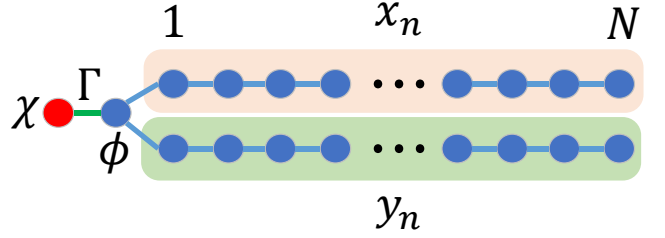

**Figure 2**. Linear chain coupled to a single extra site.

The numerical diagonalization of the H matrix of this system shows the existence of a localized state $\varphi$ with negative eigenvalue $\epsilon_0$. The latter represents, using a quantum mechanical terminology, the ground state of the system and it defines the phase transition features.

The eigenvector $\varphi$ and the associated eigenvalue $\epsilon_0$ can be analytically determined. Following the general theory recalled before, the extremal condition for the order parameter profile can be written in the following form:

$$2\mathcal{D}\chi - \Gamma\phi = \epsilon_0 \chi \\ 2\mathcal{D}\phi - \mathcal{D}(x_1 + y_1) - \Gamma\chi = \epsilon_0 \phi \\ 2\mathcal{D}f_n - \mathcal{D}(f_{n+1} + f_{n-1}) = \epsilon_0 f_n \tag{21}$$

where $f_n \in \{x_n, y_n\}$, $n \in \{1, \cdots, N\}$, $f_0 = \phi$ and $f_{N+1} = 0$. First, we observe that the $\chi$ variable can be eliminated by using the first relation, which implies:

$$\chi = \frac{\Gamma\phi}{2\mathcal{D} - \epsilon_0}.$$

Now, we are searching for a solution whose structure is $f_n = \phi\rho^n$ with $n \in \{0, \cdots, N\}$ and $\rho < 1$. Under the above assumptions, considering that $x_1 = y_1 = \phi\rho$, we obtain the following relations:

$$\rho^2 + z\rho + 1 = 0 \\ z^2 + 2\rho z - \gamma^2 = 0 \tag{22}$$

with $z = \epsilon_0 \mathcal{D}^{-1} - 2 \neq 0$ and $\gamma = \Gamma\mathcal{D}^{-1}$. Solving Equation (22), we get:

$$\rho = \sqrt{\sqrt{1 + \frac{\gamma^4}{4}} - \frac{\gamma^2}{2}}. \\ \frac{\epsilon_0}{\mathcal{D}} = 2 - \sqrt{2 + \sqrt{4 + \gamma^4}} \tag{23}$$

Equation (23) provides a complete solution of the problem once the onsite order parameter $\phi$, which parametrizes the solution, has been determined.

Equation (23) also shows that, for increasing $\gamma$ values, $\rho \to 0$, implying that a localized solution for the order parameter profile exists. When the $\gamma = 1$ case is considered, we obtain:

$$\rho = \sqrt{\frac{\sqrt{5}-1}{2}} \approx 0.786$$
$$\frac{\epsilon_0}{\mathcal{D}} = 2 - \sqrt{2+\sqrt{5}} \approx -0.058$$

which is in agreement with the numerical diagonalization of $H$. Using Equation (11), we can determine the transition temperature:

$$\mathcal{T}_c = T_c\left[1 + \frac{\mathcal{D}}{\alpha T_c}\left(\sqrt{2+\sqrt{5}}-2\right)\right],$$

the latter presenting the important enhancement property given by the relation $\mathcal{T}_c > T_c$. The above example shows the existence of a conceptual link between the order parameter localization and the critical temperature enhancement. In this subsection, we have presented the simplest case in which an exponential localization takes place. In particular, we have found that the order parameter is peaked on the island with the highest connectivity, which is consistent with our previous observations. This island behaves like a topological defect for the network. The topological defect forms bonding and antibonding states. The bonding state behaves like a potential well. The potential well, on its turn, is able to localize the order parameter following the same phenomenology of defect states of the Schrödinger problem on a lattice. Thus, the network topology can induce effective potentials, which are able to localize the superconducting order parameter. The localization of the order parameter is accompanied by a critical temperature enhancement.

The localization degree of the order parameter can be controlled by designing *ad hoc* network topologies. In particular, networks with highly connected islands can be designed. Star graph networks are a relevant example of this idea, which will be discussed in the following subsection.

### 3.3 Star graph geometries

Let us study the star network depicted in Figure 3. It consists of $p$ legs connected to an extra site, the latter representing the island with higher connectivity. Proceeding according to our general procedure, we have to determine the ground state $\varphi$ and the corresponding eigenvalue $\epsilon_0$ of the $H$ matrix, which retains information on the network connectivity. Let us assume homogeneous network parameters so that the onsite energy $2\mathcal{D}$ and the Josephson coupling constant $-\mathcal{D}$ do not depend on the considered island along the network. The extremal condition of the free energy allows us to write the following relations:

$$\begin{aligned}2\mathcal{D}\psi_n^{(\nu)} - \mathcal{D}\big(\psi_{n+1}^{(\nu)} + \psi_{n-1}^{(\nu)}\big) &= \epsilon_0\, \psi_n^{(\nu)}\\ 2\mathcal{D}\psi_N^{(\nu)} - \mathcal{D}\big(\phi + \psi_{N-1}^{(\nu)}\big) &= \epsilon_0\, \psi_N^{(\nu)}\\ 2\mathcal{D}\phi - \mathcal{D}\sum_{\nu=1}^{p}\psi_N^{(\nu)} &= \epsilon_0\, \phi\end{aligned} \quad (24)$$

where $\nu \in \{1,\cdots,p\}$ labels the system legs, $\psi_n^{(\nu)}$ represents the leg order parameter and $\phi$ is the order parameter of the island connected to all the system legs. Equations (24) can be solved by using the following ansatz for the order parameter profile

$$\begin{aligned}\psi_n^{(\nu)} &= \phi\rho^{N+1-n}\\ \rho &< 1\\ n &\in \{1,\cdots,N\}\end{aligned} \quad (25)$$

the latter being justified by the numerical diagonalization of $H$. Using Equation (25) in (24), we get the following relations:

$$\begin{aligned}\frac{\epsilon_0}{\mathcal{D}} &= 2 - p\rho\\ \rho^2 + \left(\frac{\epsilon_0}{\mathcal{D}}-2\right)\rho + 1 &= 0\end{aligned}. \quad (26)$$

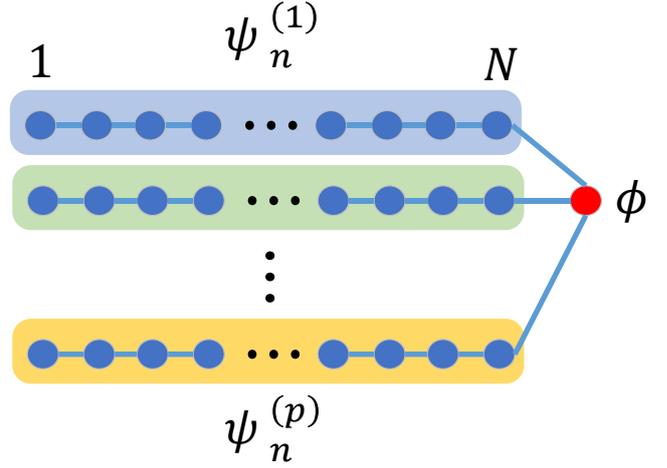

**Figure 3**. Star network with *p* legs.

Equations (26) admit the following solution:

$$\rho = \frac{1}{\sqrt{p-1}}$$
$$\frac{\epsilon_0}{\mathcal{D}} = 2 - \frac{p}{\sqrt{p-1}},$$

showing that a localized solution exists only when $p > 2$. The degree of localization increases as the number of legs in the network increases. Moreover, the system critical temperature takes the following form:

$$\mathcal{T}_c = T_c \left[1 + \frac{\mathcal{D}}{\alpha T_c}\left(\frac{p}{\sqrt{p-1}} - 2\right)\right], \quad (27)$$

which is enhanced compared to $T_c$.

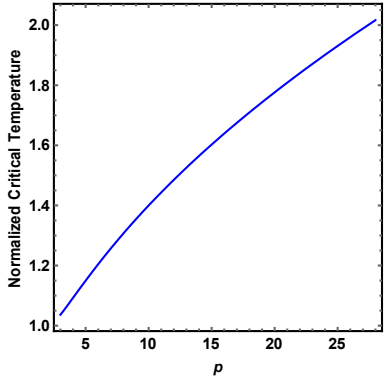

**Figure 4.** Normalized critical temperature $\mathcal{T}_c/T_c$ as a function of the number of legs $p$ obtained by using Equation (27) with $\mathcal{D}(\alpha T_c)^{-1} = 0.3$.

Equation (27) suggests that synthetic systems with high transition temperature can be obtained starting from a convenient network of low transition temperature islands. In this respect, a fascinating question is whether the above findings are promising to achieve metamaterials showing room temperature superconductivity. Figure 4 provides a partial answer to this question. Indeed, the analysis of Figure 4 suggests that a critical temperature doubling can be achieved in star graphs. In view of the relevance of this observation, it would be important to experimentally verify the validity of these deductions by performing systematic studies.

## 4. Discussion

We have demonstrated that the phase boundary of a superconducting transition can be determined by neglecting quartic terms of the free energy. The relevance of such terms has been previously discussed.

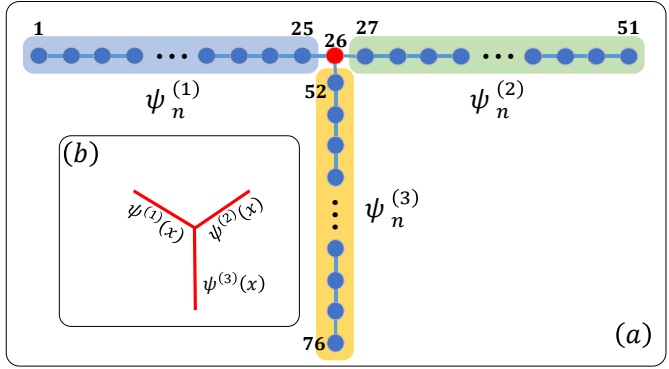

**Figure 5.** (a) Three legs star graph ($p = 3$). Islands labelled by $n \in \{1, \cdots, 25\}$ belong to the $\nu = 1$ leg; islands labelled by $n \in \{27, \cdots, 51\}$ belong to the $\nu = 2$ leg; islands labelled by $n \in \{52, \cdots, 76\}$ belong to the $\nu = 3$ leg. The island at the junction between the three legs is labelled by $n = 26$, which is coupled with the island labelled by $n = 52$. (b) Continuous version of the three legs model.

In this section, we provide further evidences of the relevance of these terms by studying the phase transition of the three legs system depicted in Figure 5a. Before discussing this topic, it is worth mentioning that the continuous version of the three legs star graph (see Figure 5b) can be studied by using the de Gennes-Alexander model. Interestingly, when a continuous formulation is considered, exponentially localized states are not supported. For this reason, we conclude that exponential localization of the superconducting order parameter is a peculiarity of the granular theory described in this work.

We have numerically studied the system depicted in Figure 5a by retaining quartic terms of the free energy and by fixing $\mathcal{D}(\alpha T_c)^{-1} = 0.1$. The results of this analysis are shown in Figure 6.

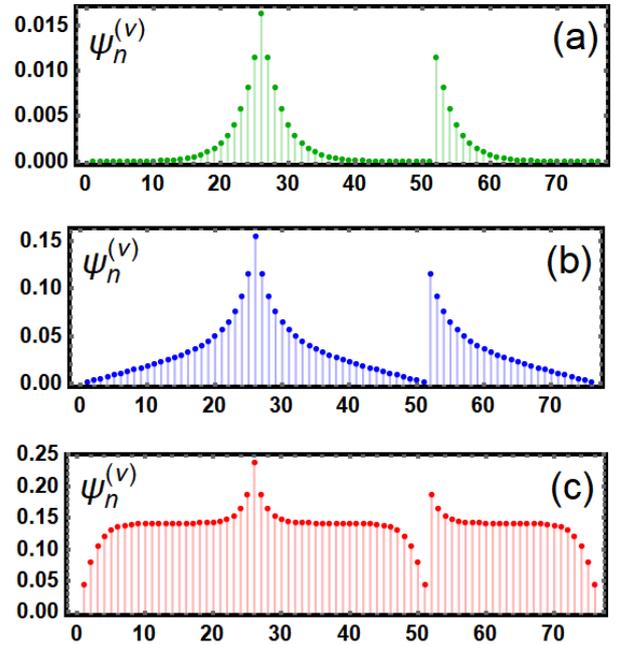

**Figure 6.** Order parameter $\psi_n^{(\nu)}$ for three legs star graph ($p = 3$) shown in Figure 5a. The order parameter profile has been computed by taking into account nonlinear terms and by fixing $\mathcal{D}(\alpha T_c)^{-1} = 0.1$. The system critical temperature is $\mathcal{T}_c/T_c \approx 1.01213$. Different panels have been obtained by fixing different temperature values. (a) $T = 1.012 \, T_c$; (b) $T = 1.0 \, T_c$; (c) $T = 0.98 \, T_c$.

Numerical simulations show a phase transition temperature given by $\mathcal{T}_c/T_c \approx 1.01213$, which perfectly agrees with Equation (27) with $p = 3$. For $T < \mathcal{T}_c$ but in close vicinity of the transition temperature (Figure 6a), the order parameter profile is localized on the junction island and does not present deviation from the ground state $\varphi$ of the $H$ matrix. Under this condition, the superconducting state nucleates at the island with higher connectivity. When the temperature is lowered (Figure 6b), the superconducting state spreads and even distant sites from the initial nucleation centre are reached. The diffusion of the order parameter brings the whole system in the superconducting phase. A further temperature lowering (Figure 6c) stabilizes the superconducting phase. Under this condition, the order parameter tends to be uniform along the system. A suppression of the order parameter is observed at the system boundaries. The order parameter enhancement remains observable at the junction island. The above results are in qualitative agreement with Equation (16).

## 5. Conclusion

We have formulated a de Gennes-Alexander micronetworks theory adapted to describe granular systems. The theory is well suited to describe arbitrary connections of superconducting islands interacting via the Josephson coupling. These systems can be of natural or synthetic origin and their phase transitions present similarities with the insulator-superconductor phase changes. We have demonstrated that at the transition temperature $\mathcal{T}_c$ the order parameter profile is exponentially localized on islands with higher connectivity. Exponential localization of the superconducting order parameter implies that the transition temperature of the system is enhanced compared to the one of the superconducting material used to realize the system islands (i.e., $\mathcal{T}_c > T_c$). This observation suggests an alternative path towards the realization of room temperature superconducting systems, being the latter idea worth to be experimentally tested.

As a relevant application of the theory, we have carefully studied star networks, which have been experimentally realized using arrays of Josephson junctions. In recent literature, the superconducting properties of these systems have been considered as a fingerprint of the Bose-Einstein condensation of Cooper pairs. Here we have demonstrated that an alternative theoretical interpretation of the experimental findings seems to be possible.

Furthermore, in view of the analogies of the present formulation with the Landau theory of the paramagnetic-ferromagnetic transition, the question arises whether the order parameter focalization phenomenon described in this work is also detectable in magnetic nanostructures.

## Acknowledgements

The author acknowledges inspiring discussions with Matteo Cirillo and Sergio Pagano. Fruitful discussions with Mario Salerno, Federico Corberi and Roberto De Luca are also acknowledged. Roberto De Luca is also acknowledged for careful reading of the paper.

## References

[1] Burioni R, Cassi D, Meccoli I, Rasetti M, Regina S, Sodano P and Vezzani A 2000 Bose-Einstein condensation in inhomogeneous Josephson arrays *Europhysics Letters (EPL)* **52** 251-256

[2] Burioni R, Cassi D, Rasetti M, Sodano P and Vezzani A 2001 Bose-Einstein condensation on inhomogeneous complex networks *Journal of Physics B: Atomic, Molecular and Optical Physics* **34** 4697-4710

[3] Buonsante P, Burioni R, Cassi D and Vezzani A 2002 Bose-Einstein condensation on inhomogeneous networks: Mesoscopic aspects versus thermodynamic limit *Physical Review B* **66** 094207

[4] Silvestrini P, Russo R, Corato V, Ruggiero B, Granata C, Rombetto S, Russo M, Cirillo M, Trombettoni A and Sodano P 2007 Topology-induced critical current enhancement in Josephson networks *Physics Letters A* **370** 499–503

[5] Lorenzo M, Lucci M, Merlo V, Ottaviani I, Salvato M, Cirillo M, Müller F, Weimann T, Castellano M G, Chiarello F and Torrioli G 2014 On Bose–Einstein condensation in Josephson junctions star graph arrays *Physics Letters A* **378** 655-658

[6] Ottaviani I, Lucci M, Menditto R, Merlo V, Salvato M, Cirillo M, Müller F, Weimann T, Castellano M G, Chiarello F, Torrioli G and Russo R 2014 Characterization of anomalous pair currents in Josephson junction networks *Journal of Physics: Condensed Matter* **26** 215701

[7] Lucci M, Cassi D, Merlo V, Russo R, Salina G and Cirillo M 2020 Conditioning of Superconductive Properties in Graph-Shaped Reticles *Scientific Reports* **10** 10222

[8] Anderson P W 1987 The Resonating Valence Bond State in $La_2CuO_4$ and Superconductivity *Science* **235** 1196-1198

[9] Grassie A D C and Green D B 1970 The transitions to superconductivity of disordered films *Journal of Physics C: Solid State Physics* **3** 1575-1586

[10] Abeles B, Cohen R W and Cullen G W 1966 Enhancement of Superconductivity in Metal Films *Physical Review Letters* **17** 632-634

[11] Cohen R W and Abeles B 1968 Superconductivity in Granular Aluminum Films *Physical Review* **168** 444

[12] Deutscher G, Fenichel H, Gershenson M, Grünbaum E and Ovadyahu Z 1973 Transition to zero dimensionality in granular aluminum superconducting films *Journal of Low Temperature Physics* **10** 231-243

[13] Ziemann P, Heim G and Buckel W 1978 Oxygen content and oxide barrier thickness in granular aluminum films *Solid State Communications* **27** 1131-1135

[14] Berger J and Rubinstein J 2000 *Connectivity and superconductivity* (Berlin; New York: Springer)

[15] Simonin J 1986 Surface term in the superconductive Ginzburg-Landau free energy: Application to thin films *Phys. Rev. B* **33** 7830(R)

[16] Sacco C, Galdi A, Romeo F, Coppola N, Orgiani P, Wei H I, Goodge B H, Kourkoutis L F, Shen K, Schlom D G and Maritato L 2019 Carrier confinement effects observed in the normal-state electrical transport of electron-doped cuprate trilayers *Journal of Physics D: Applied Physics* **52** 135303

[17] de Gennes P-G 1999 *Superconductivity of metals and alloys* (Reading, Mass: Advanced Book Program, Perseus Books)